\documentclass[a4paper, amsfonts, amssymb, amsmath, reprint, showkeys, twoside, superscriptaddress,citeautoscript]{revtex4-1}

\UseRawInputEncoding
\usepackage[utf8]{inputenc}
\usepackage[T1]{fontenc}
\inputencoding{utf8}
\usepackage{xcolor}
\usepackage{adjustbox}
\usepackage[colorlinks,allcolors=black,citecolor=blue,urlcolor=blue]{hyperref}
\usepackage{todonotes}
\usepackage{dirtytalk}
\usepackage{bm}
\usepackage{amsmath}
\usepackage{comment}
\usepackage[capitalise]{cleveref}

\setcitestyle{super}

\newcommand{\rev}[1]{{\color{black} #1}}

\usepackage{amsmath}

\DeclareMathOperator*{\argmin}{arg\,min}
\newcommand{\invcm}{$\text{cm}^{-1}$}
\newcommand{\qc}{\mathbf{q}_\textrm{c}}
\newcommand{\fc}{\mathbf{f}_\textrm{c}}
\newcommand{\dint}{\textrm{d}}
\newcommand{\kb}{k_{\mathrm{B}}}
\newcommand{\Te}{\textrm{T}_{\mathrm{e}}}
\newcommand{\TePIGS}{$\textrm{T}_{\mathrm{e}}$~PIGS}
\newcommand{\average}[2]{\left\langle #1 \right\rangle_{#2}}

\begin{document}

\title{Quantum dynamics using path integral coarse-graining}
\author{F\'elix Musil}
\affiliation{Department of Physics, Freie Universit\"at Berlin, Arnimallee 12, 14195 Berlin, Germany}

\author{Iryna Zaporozhets}
\affiliation{Department of Physics, Freie Universit\"at Berlin, Arnimallee 12, 14195 Berlin, Germany}
\affiliation{Department of Chemistry, Rice University, Houston, Texas 77005, United States}
\affiliation{Center for Theoretical Biological Physics, Rice University, Houston, Texas 77005, United States}

\author{Frank No\'e}
\affiliation{Department of Physics, Freie Universit\"at Berlin, Arnimallee 12, 14195 Berlin, Germany}
\affiliation{Center for Theoretical Biological Physics, Rice University, Houston, Texas 77005, United States}
\affiliation{Department of Chemistry, Rice University, Houston, Texas 77005, United States}
\affiliation{Microsoft Research, Cambridge, UK}

\author{Cecilia Clementi}
\email{cecilia.clementi@fu-berlin.de}
\affiliation{Department of Physics, Freie Universit\"at Berlin, Arnimallee 12, 14195 Berlin, Germany}
\affiliation{Center for Theoretical Biological Physics, Rice University, Houston, Texas 77005, United States}
\affiliation{Department of Chemistry, Rice University, Houston, Texas 77005, United States}

\author{Venkat Kapil}
\email{vk380@cam.ac.uk}
\affiliation{Yusuf Hamied Department of Chemistry,  University of Cambridge,  Lensfield Road,  Cambridge,  CB2 1EW,UK}

\begin{abstract}
Vibrational spectra of condensed and gas-phase systems containing light nuclei are influenced by their quantum-mechanical behaviour. 
The quantum dynamics of light nuclei can be approximated by the imaginary time path integral (PI) formulation, but still at a large computational cost that increases sharply with decreasing temperature. 
By leveraging advances in machine-learned  coarse-graining, we develop a PI method with the reduced computational cost of a classical simulation. 
We also propose a simple temperature elevation scheme to significantly attenuate the artefacts of standard PI approaches and also eliminate the unfavourable temperature scaling of the computational cost.
We illustrate the approach, by calculating  vibrational spectra using standard models of water molecules and bulk water, demonstrating significant computational savings and dramatically improved accuracy compared to more expensive reference approaches. 
We believe that our simple, efficient and accurate method could enable routine calculations of vibrational spectra including nuclear quantum effects for a wide range of molecular systems. 
\end{abstract}

\maketitle

Predictive simulations of thermodynamic and time-dependent properties of condensed and gas-phase systems lay the foundations of computational materials design and discovery~\cite{houk_holy_2017}.
Accurate modeling of many systems, such as those containing light nuclei like H, C, N, and O, must account for their quantum-mechanical behavior to include the zero-point motion of collective modes~\cite{pereyaslavets_importance_2018} and the tunneling of the system across classically inaccessible barriers~\cite{benoit_tunnelling_1998}.
Phenomena emerging from the quantum dynamics of light nuclei are ubiquitous in chemistry and material science of molecular systems, for instance, the relative diffusion of $\mathrm{H}_2$ in clathrate hydrates~\cite{cendagorta_enhanced_2021} and kinetic isotope effects in porous organic crystals~\cite{liu_barely_2019}, proton-transfer rates in molecular switches~\cite{litman_elucidating_2019}, the red-shift in the IR spectra of O--H stretch mode in ice~\cite{burnham_origin_2006} and the characterization of (bio-)molecular systems using vibrational spectroscopy~\cite{rossi_stability_2015}.  \\

Unfortunately, the exact description of quantum dynamics -- requiring the solution of Schrödinger's equation -- is possible only for the smallest of systems, such as molecules containing a few atoms~\cite{li_vibrational_2001, larsson_state-resolved_2022}.
Extension to larger systems requires ``local" approximations or truncation of the interaction potential and/or an approximate solution of the many-body Schrödinger equation~\cite{wang_ab_2011, gruenbaum_robustness_2013, qu_multimode_2021}, akin to the electronic structure problem.
Alternatives that render an ``approximate but full" quantum-mechanical treatment of all degrees of freedom are based on the (semi-)classical dynamics of the system~\cite{althorpe_path-integral_2021}. 
In this context, imaginary time path integral (PI) simulations, although originally formulated for incorporating the quantum statistical effects ~\cite{markland_nuclear_2018}, are becoming increasingly popular for studying the approximate quantum dynamics of distinguishable particles~\cite{althorpe_path-integral_2021}. 
State-of-the-art PI approaches~\cite{cao_formulation_1994, craig_quantum_2004, rossi_how_2014, kapil_inexpensive_2020} neglect real time quantum coherence but include effects arising from the quantum statistical distribution of the nuclei~\cite{hele_boltzmann-conserving_2015}.
These methods give a good description of the dynamical response of condensed phase systems, for instance, vibrational spectra of bulk water~\cite{marsalek_quantum_2017} and ice~\cite{kapil_inexpensive_2020}, where quantum coherence effects last only over short times. 
On the other hand, predicting the vibrational response of molecules and clusters is still a challenge, the reasons being twofold.
Firstly, the artifacts of PI methods (arising from the neglect of real-time coherence and approximations to non-centroid Matsubara fluctuations~\cite{trenins_mean-field_2018}), such as spurious broadening, frequency shifts, \rev{and an incorrect temperature-dependence of the relative intensities of quantal modes}~\cite{benson_matsubara_2021, ple_anharmonic_2021} become worse with a reduction in temperature~\cite{witt_applicability_2009}.
Secondly, the computational cost of PI methods increases steeply with inverse temperature~\cite{uhl_accelerated_2016} making them expensive to obtain a direct comparison with experiments.
In the last few decades, many new approaches have aimed at individually improving the accuracy~\cite{rossi_fine_2017, willatt_approximating_2018, liu_path_2014, trenins_path-integral_2019} or efficiency~\cite{kapil_accurate_2016, kapil_inexpensive_2020, shepherd_efficient_2021, fletcher_fast_2021} of PI dynamical methods.
These studies highlight the urgency for an accurate, efficient, and generally-applicable method that can treat the quantum dynamics of molecules, clusters, and bulk systems at the same footing, and aid in modeling their vibrational response. \\

\begin{figure*}[!t]
    \centering
    \includegraphics[width=0.96\textwidth]{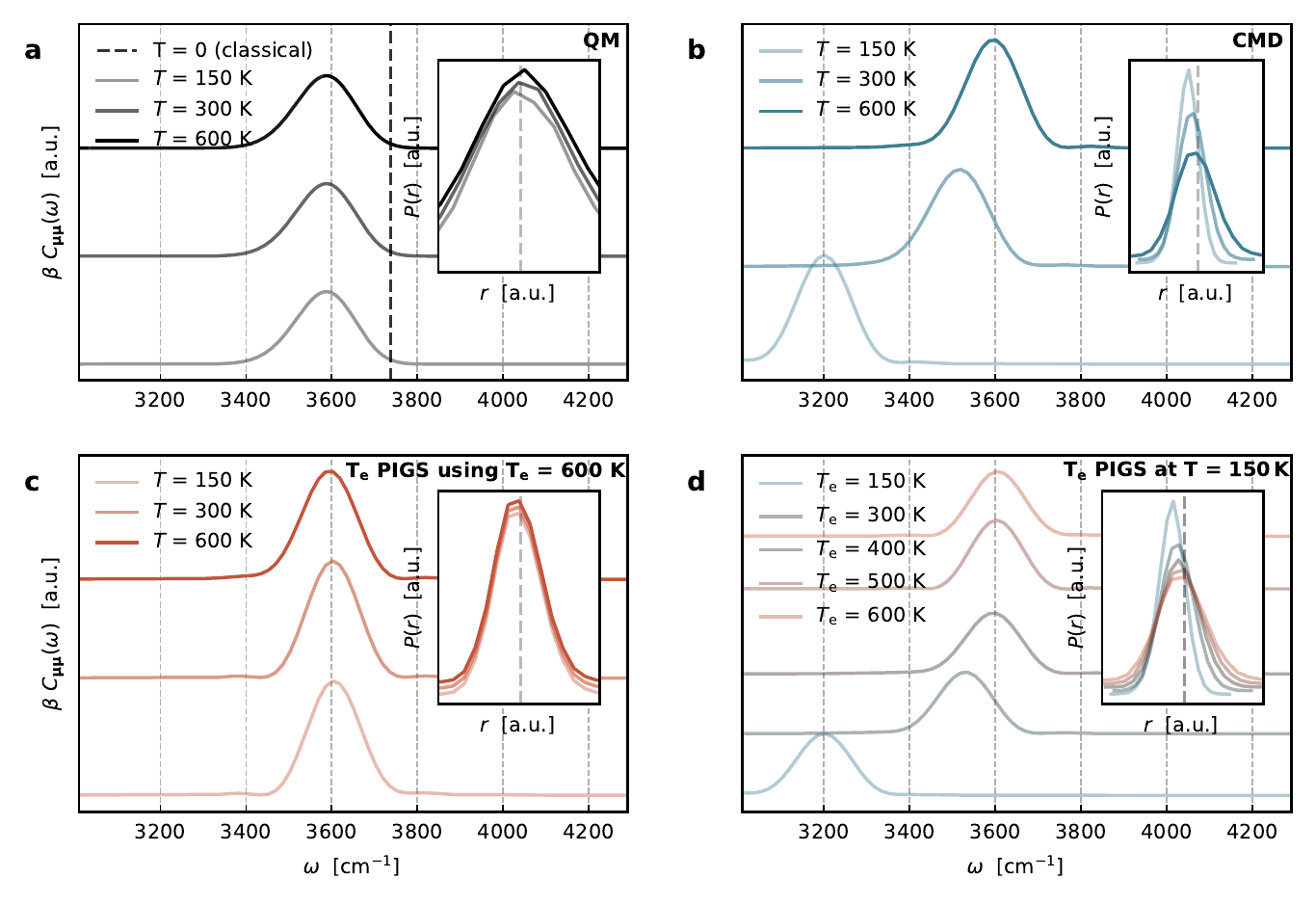}
    \caption{\scriptsize\textbf{Temperature dependent IR spectrum of an O--H bond described by a Morse oscillator.} The IR spectrum of a 2D Morse oscillator mimicking an O--H at 600\,K, 300\,K, 150\,K calculated by (a) solving the Schr\"odingers equation numerically (QM)~\cite{trenins_path-integral_2019}, and using (b) centroid molecular dynamics (CMD), and (c) \TePIGS{} using $\Te{}$ = 600\,K. Panel (d) shows the dependence of the \TePIGS{} IR spectrum on the temperature of the centroid free energy surface, ($\Te{}$). Insets show the temperature-dependent radial probability densities calculated using the respective methods and the dashed grey vertical line is the mode of quantum-mechanical distribution. Note that the range of $y$ axes is not kept the same to aid clarity.  
    \label{fig:morse}}
\end{figure*}

In this work, we combine PI and coarse-graining methods via machine learning to render the calculation of quantum vibrational spectra accurate and computationally affordable and demonstrate its capabilities on paradigmatic aqueous systems. 
Our approach builds upon the state-of-the-art centroid molecular dynamics~\cite{cao_new_1993, cao_formulation_1994} (CMD) approach, which time-evolves the system \textit{classically} on the free energy surface (FES) of the centroid of the imaginary time path -- a modified PES that includes nuclear quantum effects.
Its two key features, a temperature elevation ($\Te{}$) \textit{ansatz} and path integral coarse-graining simulations (PIGS), ensure accuracy and computational efficiency.
On one hand, the $\Te{}$ \textit{ansatz} alleviate the spurious redshift in CMD, leading to an improvement in the accuracy of vibrational spectra over state-of-the-art methods.
On the other hand, with  PIGS we machine learn the centroid FES in a general manner, i.e., without making prior assumptions about the functional form of the system's PES, and use it to evolve the system classically.
This method, referred to as \TePIGS{}, enables the calculation of the IR spectra of aqueous systems including nuclear quantum effects, in excellent agreement with numerically exact or more expensive reference methods.
Furthermore, we demonstrate that our approach is transferable across phases and temperatures allowing modeling of vibrational spectra at \textit{cryogenic} temperatures at orders of magnitude lower computational cost using classical MD. \\

We first discuss CMD and introduce \TePIGS{} in the context of a simple yet realistic anharmonic system that highlights the deficiencies of PI methods at low temperatures: a 2D radial Morse oscillator mimicking an O--H bond, described by the Hamiltonian,

\begin{align}
    \hat{H} = (2 \mu)^{-1} \left[\hat{p}_x^2 + \hat{p}_y^2\right] + D \left[1 - e^{-\alpha \left(\sqrt{\hat{q}_x^2 + \hat{q}_y^2} - r_0\right)}\right]^2, 
\label{eq:2d_morse_Ham}
\end{align}

where the parameters $\mu$, $D$, $\alpha$, and $r_0$ are defined in section I.A of the supporting information (SI). 
\rev{For simplicity, we ignore coupling between the rotations and vibrations of the system, making this a simple but physically relevant one-dimensional problem that highlights limitations of PI based dynamical methods.} 
The O--H bond has a large zero point energy $E_0 \approx 1843$ \invcm{} $\approx 2652$\,K and a $0\to1$ transition energy $E_1 - E_0 \approx 3568$ \invcm{} $\approx  5134$\,K (see section I.A of the SI for more details), meaning that even at a temperature of 600\,K the system resides almost exclusively in its ground state, yet probes anharmonic regions of the PES due to zero-point motion. 
As shown in Fig.~\ref{fig:morse}(a), this results in a temperature-independent line position of the IR spectrum, at least up to 600\,K, but also a red-shift of around 200 \invcm{} with respect to the classical spectrum due to quantum nuclear motion. \\
\begin{figure}[!t]
    \centering
    \includegraphics[width=0.48\textwidth]{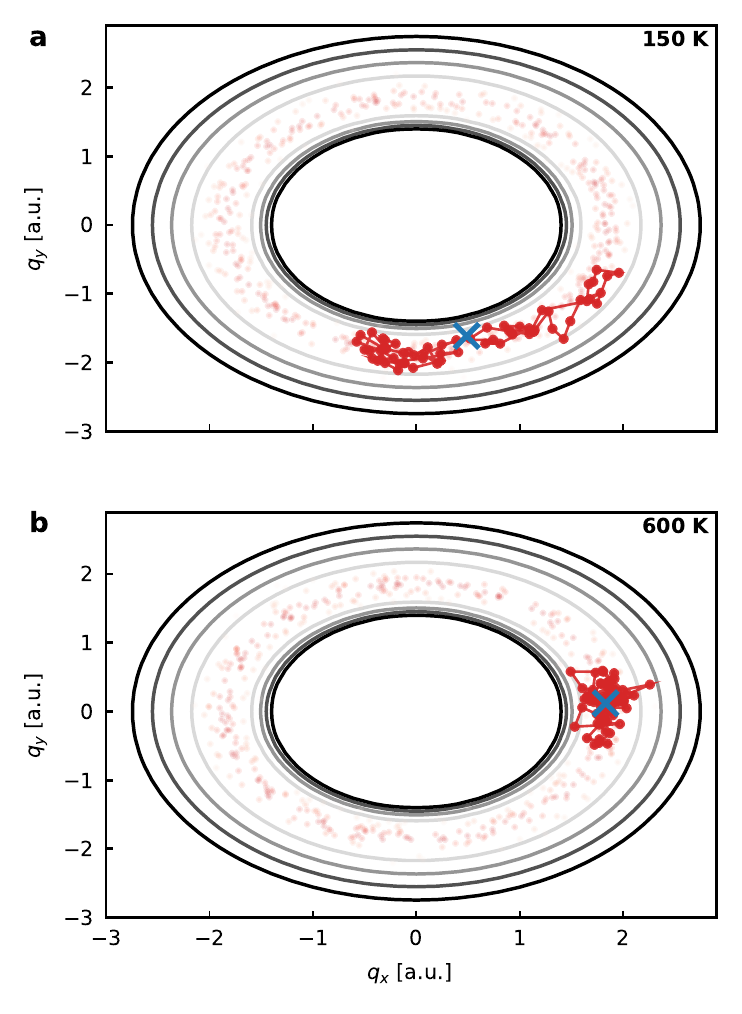}
    \caption{\scriptsize\textbf{The curvature problem in centroid molecular dynamics} Contour plots of a 2D Morse radial potential up to half the dissociation energy (black to grey solid lines), the probability density of the system (red to white indicating high to low), and a snapshot of the ring polymer (red) and its centroid (blue) at (a) 150\, K and (b) 600\,K. 
    \label{fig:curvature_problem}}
\end{figure}

The CMD approach~\cite{cao_formulation_1994} is based on the imaginary time path integral isomorphism~\cite{chandler_exploiting_1981} between the thermodynamics of a quantum system at inverse temperature $\beta$ and a classical ring polymer made of $P$ replicas of the system at $\beta_P = \beta / P$, 

\begin{align}
    Z = \text{Tr}[e^{-\beta \hat{H}}] \propto \lim_{P\to\infty} \int \dint{}\mathbf{q}~ e^{-\beta_P \left[\sum_j U(\mathbf{q}^{(j)}) + U^{\textrm{spr}}(\mathbf{q})\right]},
    \label{eq:isomorphism}
\end{align}

where $\mathbf{q} \equiv \{\mathbf{q}^{(1)},\dots,\mathbf{q}^{(P)}\}$ is a shorthand for positions of the $P$ replicas of the system with $\mathbf{q}^{(j + P)} \equiv \mathbf{q}^{(j)}$ implied, $U(\mathbf{q})$ defines the classical PES, and $U^{\textrm{spr}}(\mathbf{q})$ is a temperature dependent spring term~\cite{chandler_exploiting_1981} that connects consecutive replicas of the system.
Within CMD, the system is time evolved \textit{classically} on $U^{\mathrm{CMD}}(\qc; \beta)$,  defined as the free energy surface of the centroid of the imaginary time path (modulo a constant) at $\beta$,

\begin{align}
 U^{\mathrm{CMD}}(\qc; \beta) = -\beta^{-1} \log \left\langle ~\delta\Big(\frac{1}{P}\sum_{j=1}^{P}\mathbf{q}^{(j)}  - \mathbf{q}_c\Big)\right\rangle_{
\beta},
\label{eq:cmd_pmf}
\end{align}

where $\sum_{j=1}^{P}\mathbf{q}^{(j)}/ P$ is the centroid of the ring polymer and $\langle\cdot\rangle$ an average over the path integral Hamiltonian in Eq.~\ref{eq:isomorphism}. 
The thermodynamic force acting on the centroid can be calculated \textit{on the fly} from a constrained PI simulation at each CMD step~\cite{cao_formulation_1994}

\begin{align}
& \mathbf{f}^{\mathrm{CMD}}(\qc; \beta) = -\nabla U^{\mathrm{CMD}}(\qc; \beta) \nonumber \\
& = \left\langle \fc ~\delta\Big(\frac{1}{P}\sum_{j=1}^{P}\mathbf{q}^{(j)} - \qc\Big) \right\rangle_{\beta}, 
\label{eq:cmd_force}
\end{align}

where, $\fc = \sum_{j=1}^{P}\mathbf{f}^{(j)} / P$, and  $\mathbf{f}^{(j)}$ is the physical force acting on the $j$-th replica.
Alternatively, the centroid can be evolved on its FES within a PI simulation in a partially adiabatic manner~\cite{hone_comparative_2006} by decoupling the centroid from the rest of the system. 
Dynamical properties, such as the IR spectrum, can be easily calculated via classical time correlation functions (TCFs) on the basis of the centroid trajectory, akin to classical MD. \\

As shown in Fig.~\ref{fig:morse}(b), the IR spectrum of the O--H bond computed with CMD is in excellent agreement with the numerically exact result at 600\,K. 
Unfortunately, at lower temperatures, the system experiences a spurious red shift that gets worse as the temperature is reduced. 
This well-known artifact is referred to as the ``curvature problem"~\cite{witt_applicability_2009} and arises in cases where the ring polymer has a shape such that its centroid lies outside the ring (see Fig.~\ref{fig:curvature_problem}) and therefore does not represent the quantum-mechanical probability density of the system (as obtained by the replicas).
Interestingly, the curvature problem is purely a structural artifact, and, as shown in the inset of Fig.~\ref{fig:morse}(b), it can be diagnosed from PI trajectories: the misalignment of the distribution of the centroid at 300\,K and 150\,K to the (physical) probability density obtained from the replicas. 
As mentioned above, another issue with CMD, and more generally with PI methods, is that the required number of replicas scales inversely with temperature and the maximum physical frequency of the system, $P \sim \beta \hbar \omega_{\mathrm{max}}$~\cite{markland_efficient_2008}.
Therefore its computational cost increases steeply as the temperature is reduced.
These challenges prevent investigations of systems at cryogenic temperatures where most experimental data is available. \\

Recently, ~\citet{trenins_path-integral_2019} have proposed a \textit{quasi} centroid molecular dynamics (QCMD) scheme, that evolves the system on the FES of an \textit{ad hoc} curvilinear function of replica positions -- a \textit{quasi} centroid that does not ``fall out" of the hull of the path integral. 
A careful selection of this function doesn't hamper quantum Boltzmann statistics and results in a compact PI ring polymer that alleviates the curvature problem.
This results in an excellent agreement of the vibrational spectrum of a molecule of water, liquid water, and proton disordered hexagonal ice~\cite{benson_which_2019} with (numerically exact) reference methods. 
More recently, ~\citet{fletcher_fast_2021} have proposed an efficient \textit{fast} QCMD scheme that avoids the costly \textit{on the fly} quasi-centroid forces by precomputing an analytic FES on the basis of a PI trajectory. 
This reduces the cost of predicting an accurate vibrational spectrum to that of classical MD, resulting in an improvement over standard PI methods for the calculation of vibrational spectra of exemplary molecular systems.~\cite{fletcher_fast_2021} 
Unfortunately, the computational cost of these approaches still grows unfavorably with temperature due to the need for a low-temperature PI trajectory for fitting the FES.
Moreover, an extension to general systems requires a general/universal procedure to fit the FES of the centroid and careful knowledge of appropriate curvilinear coordinates that do not suppress sampling of the physical regions of the configurational space~\cite{haggard_testing_2021}.
Nonetheless, the advancements brought forth by (\textit{fast}) QCMD~\cite{trenins_path-integral_2019, fletcher_fast_2021} suggest that further improvements in CMD provide a promising route towards the accurate and efficient calculation of vibrational spectra. \\

Taking inspiration from \textit{fast} implementations of CMD~\cite{Hone2005, paesani_accurate_2006} we propose path integral coarse-grained simulations (PIGS) that perform MD on a modified PES -- including quantum nuclear effects.
In particular, we leverage the recent developments in the definition of coarse-grained machine learning potentials~\cite{Wang2019,husic2020coarse,wang2020ensemble} and of high-order correlation functions~\cite{Musil2021rev,Unke2021rev} to obtain this modified potential by coarse-graining the imaginary time path integral to that of a classical system.
In this work, we use PIGS to accelerate CMD in a general manner, i.e. it is applicable to systems exhibiting a wide range of interparticle interactions. 
To obtain the centroid FES at $\beta_\mathrm{e}$, we use the force matching method~\cite{Izvekov2005, Noid2008} that is typically used to build thermodynamically consistent bottom-up coarse-grained models for macromolecular systems, i.e. the coarse-grained model reproduces the thermodynamic properties of the all-atom system projected onto the coarse-grained coordinates.
It has been shown~\cite{Noid2008} that \cref{eq:cmd_pmf} can be recast as a variational principle: $U^{\mathrm{CMD}}(\mathbf{q}_c; \beta)$ corresponds to the minimum of the force matching functional:
\begin{equation}
     U^{\mathrm{CMD}}(\mathbf{q}_c; \beta) = \argmin_{G \in C(\mathbb{R}^{n})} \average{\Big\|\mathbf{f}_c  + \nabla G(\mathbf{q}_c) \Big\|^2 }{\beta},
\label{eq:force_matching}
\end{equation}
where the average is performed at inverse temperature $\beta$ with the PI centroid constrained at $\qc \in \mathbb{R}^{n}$, and the minimization is performed over the space $C(\mathbb{R}^{n})$ of all real continuous functions $G: \mathbb{R}^{n} \rightarrow \mathbb{R}$.
In this work we optimize a machine learning model as a surrogate for the potential of mean force to reproduce the thermodynamics of the centroid without having to explicitly simulate $P$ replicas of the system.
In practice, we only learn the difference between the centroid FES and the classical PES, as it is done in Ref.~\cite{Hone2005}
This keeps the amount of training data to a minimum and provides an appropriate prior in the absence of data or for collective modes that do not exhibit quantum nuclear effects.\\

The centroid potential of mean force associated with an atomic configuration is expressed as a sum of the classical PES associated with $U$, and atom-centered contributions 

\begin{equation}
\tilde{U}^{\mathrm{CMD}}(\mathbf{q}_c; \bm{\theta}) = U(\mathbf{q}_c) + \sum_{i} A_{\textrm{a}_i}(\mathbf{q}_c -\mathbf{q}_{c~i}; \bm{\theta}),
\label{eq:pmf_ansatz}
\end{equation}

where $\bm{\theta}$ is a set of model parameters, $\mathbf{q}_c \equiv \{\mathbf{q}_{c~1},\dots,\mathbf{q}_{c~N}\}$ is a shorthand for set of atomic positions of a structure, $\mathbf{q}_{c~i}$ is the position of the $i^{\textrm{th}}$ atom of species $\textrm{a}_{i}$, and the $A_{\textrm{a}_{i}}$ functions are parameterized using a machine learning model that captures the multi-body interactions emerging from the coarse-graining procedure~\cite{Wang2021}. 
We model the atomic potentials of mean force, $A_{\textrm{a}_i}$, by representing the atomic environment with the normalized SOAP powerspectrum~\cite{Bartok2013,Darby2021,Musil2021} and pass these 3-body features to a multi-layer perceptron with $3$ hidden layers of width $[400, 200, 200]$ and the hyperbolic tangent activation function. 
The parameters of the model are obtained by minimizing the force matching loss

\begin{equation}
     \left\Vert \mathbf{f}_c + \nabla\tilde{U}^{\mathrm{CMD}}(\mathbf{q}_c; \bm{\theta}) \right\Vert^2,
\label{eq:loss}
\end{equation}

over a set of centroid sample forces $\mathbf{f}_c$ and configurations $\mathbf{q}_c$ obtained from a PIMD simulation at $\beta$ performed with the \texttt{i-PI} code~\cite{kapil_i-pi_2019} (see section I.C of the SI for more details). The model and its training have been implemented in \texttt{pytorch}~\cite{pytorch} and the codes are available upon request. \\

\begin{figure*}[!t]
    \centering
    \includegraphics[width=\textwidth]{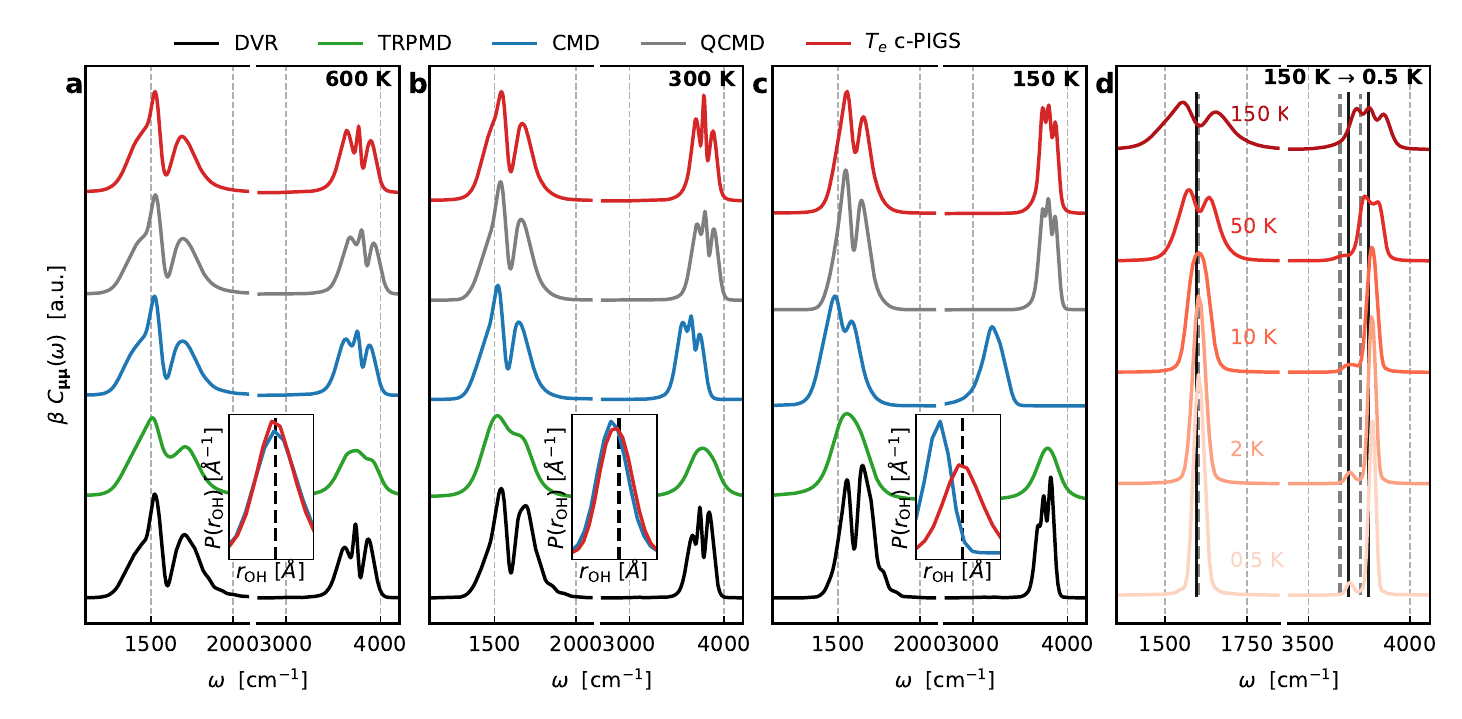}
    \caption{\scriptsize\textbf{Quantum dynamics of a water molecule.} Comparison of the vibrational spectrum of a water molecule described by the Partridge-Schwenke model\cite{partridge_determination_1997} at (a) 600\,K, (b) 300\,K, (c) 150\,K, using discrete value representation  (yielding numerically exact results) in black, the proposed $\Te{}$ PIGS approach using the centroid free energy surface calculated at 600 K in red, CMD in blue, and TRPMD in green. The insets show the probability distribution of the O--H bond length calculated from the CMD and \TePIGS{} and the black dashed line indicates the mode of the distribution obtained from PIMD -- constant across temperatures. Panel (d) displays the temperature-dependent IR spectrum of a water molecule from 600 K down to 0.5 K. The gray dashed line indicates the reference $0\rightarrow1$ transition frequency while the black line is shifted by 40~\invcm{} to reflect the results obtained with Matsubara dynamics~\cite{trenins_mean-field_2018}. }
    \label{fig:h2omol}
\end{figure*}

We propose a simple and physically motivated temperature elevation ($\Te{}$) \textit{ansatz} that alleviates the curvature problem of CMD and also eliminates the unfavorable temperature dependence associated with the computational cost of vibrational spectra.
We note that for a system in its ground state, i.e. $\hbar \omega >> \beta^{-1}$, the IR spectrum -- related by the dipole correlation function -- is only \textit{trivially} dependent on the temperature (see section II.A of the SI).
It is easy to show that after rescaling with the inverse temperature $\beta$, the Kubo-transformed time-correlation function is a constant if the system is in its ground state, i.e., $\beta \tilde{C}_{\hat{\bm{\mu}} \hat{\bm{\mu}}}(\omega; \beta) = \mathrm{constant}$.
As shown in Fig.~\ref{fig:morse}(b), CMD does not follow this behavior because of the curvature problem at low temperatures. 
\rev{To avoid this artifact, we propose a $\Te{}$ \textit{ansatz}, i.e. we rewrite a time-correlation function in terms of a CMD time-correlation function computed at an elevated temperature $\Te{}$,

\begin{align}
\tilde{C}^{\Te{}}_{\hat{\bm{\mu}} \hat{\bm{\mu}}}(\omega, \beta; \beta_\mathrm{e}) = \frac{\beta_\mathrm{e}}{\beta} \tilde{C}^{~\mathrm{CMD}}_{\hat{\bm{\mu}} \hat{\bm{\mu}}}(\omega, \beta_\mathrm{e}).
\label{eq:TEAnstaz}
\end{align}

This \textit{ansatz} is valid for any quantum system (of distinguishable particles) and can be used to improve PI based methods that exhibit diminishing performance at low temperatures.
In order to ease the calculation of quantum dynamical properties for general (high-dimensional) systems, we approximate this relation as 

\begin{align}
    \tilde{C}^{\Te{}}_{\hat{\bm{\mu}} \hat{\bm{\mu}}}(\omega, \beta; \beta_\mathrm{e}) \approx Z^{-1} \int \dint{}\mathbf{q}'~ e^{-\beta U^{\mathrm{CMD}}(\mathbf{q}'; \beta_e)}~ \bm{\mu}(0)\cdot\bm{\mu}(t),
    \label{eq:TEApprox}
\end{align} %
so that it is simply estimated by evolving the centroid at $\beta$ on a FES calculated at a high temperature $\beta_\mathrm{e}$. 
This approximate form of Eq.\ref{eq:TEAnstaz} bears advantages such as it is exact in the harmonic and the classical limit, and doesn't require any posterior rescaling of the TCF.  
These limits also suggest that Eq.\ref{eq:TEApprox} should give a good description of stiff modes, weakly coupled with the rest of the system, without perturbing the dynamics of the low frequency (classical) modes. } 
\rev{The use of an elevated temperature also improves sampling efficiency, akin to the use of a high temperature in the adiabatic free energy dynamics\cite{} to improve sampling of a high dimensional configurational space. The elevated temperature also permits the use of a small number of replicas to discretize the imaginary time path.
In this work, we perform fully adiabatic CMD by separately computing the FES at a higher temperature, however, we also plan to implement and study a partially adiabatic implementation of the $\Te{}$ \textit{ansatz} with CMD. }\\
As shown in Fig.~\ref{fig:morse}(b), the CMD IR spectrum remains in excellent agreement with the exact result at 600\,K and the radial distribution function of the centroid of the O--H bond is aligned with the physical distribution.
These observations suggest that the system doesn't exhibit the curvature problem at 600
\,K and thus we test the $\Te{}$ \textit{anstaz} for $T_\mathrm{e} = 600$\,K where $\beta_\mathrm{e} = (\kb{} T_\mathrm{e})^{-1}$.
As shown in Fig.~\ref{fig:morse}(c) and Fig.~\ref{fig:morse}(d), increasing the ``elevated temperature" progressively improves the description of the IR spectrum at 150\,K.
Moreover, using $T_\mathrm{e} = 600$\,K leads to an excellent agreement of the temperature dependent IR spectrum with the exact result.
Note that for the Morse potential there exists a wide window of suitable $T_\mathrm{e}$ (see section II.B of the SI for more details), i.e. large enough to alleviate the curvature problem but also small enough for the harmonic approximation in Eq.~\ref{eq:TEApprox} to be valid, and we observed similar features with the water molecule and bulk water.   
Fig.~\ref{fig:morse}(c) shows that the line position of the predicted IR spectra and the radial distribution of the O--H bond are largely temperature independent using $T_\mathrm{e} = 600$\,K, as expected for a system in its ground state.
Finally, the number of replicas needed to calculate $U^{\mathrm{CMD}}(\mathbf{q}'; \beta_e)$ is $P \approx \beta_e \hbar  \omega_{\mathrm{max}} < \beta \hbar  \omega_{\mathrm{max}}$ which is independent of $\beta$.
Thus, within the $T_e$ \textit{ansatz}, the cost of simulating the quantum dynamics of a system in its ground state at $\beta$ doesn't scale with temperature. \\

The workflow for computing the quantum vibrational spectrum of a system using PIGS and the $\Te{}$ \textit{ansatz} can be summarized as follow.
First, we perform short PIMD simulations exploring a range of temperatures and select the \textit{lowest} temperature $T_\mathrm{e}$ by checking for the alignment between the centroid and the physical radial distributions.
\rev{In this study, we use radial distribution functions (RDFs) to check for this alignment, however, more general systems might require using many-body correlation functions such as SOAP~\cite{Bartok2013} or ACE~\cite{Drautz2019} (of which RDFs form a subset), combined with state-of-the-art dimensionality reduction schemes to compare centroid and physical distributions.}
Then, we use \cref{eq:loss,eq:pmf_ansatz} to machine-learn the centroid FES at $\beta_\mathrm{e}$ as a sum of local atom-centered components. 
In the final step, we predict quantum dynamical properties by performing MD at the desired temperatures. 
These simple steps enable the development of an effective PES that includes quantum nuclear effects for dynamics -- an FES trained on a \textit{single} high-temperature PI trajectory -- and that is \textit{transferable} across temperature.
The computational details of all the simulations are described in sections I.B, I.D, and I.E of the SI. \\

\begin{figure*}[!t]
    \centering
    \includegraphics[width=\textwidth]{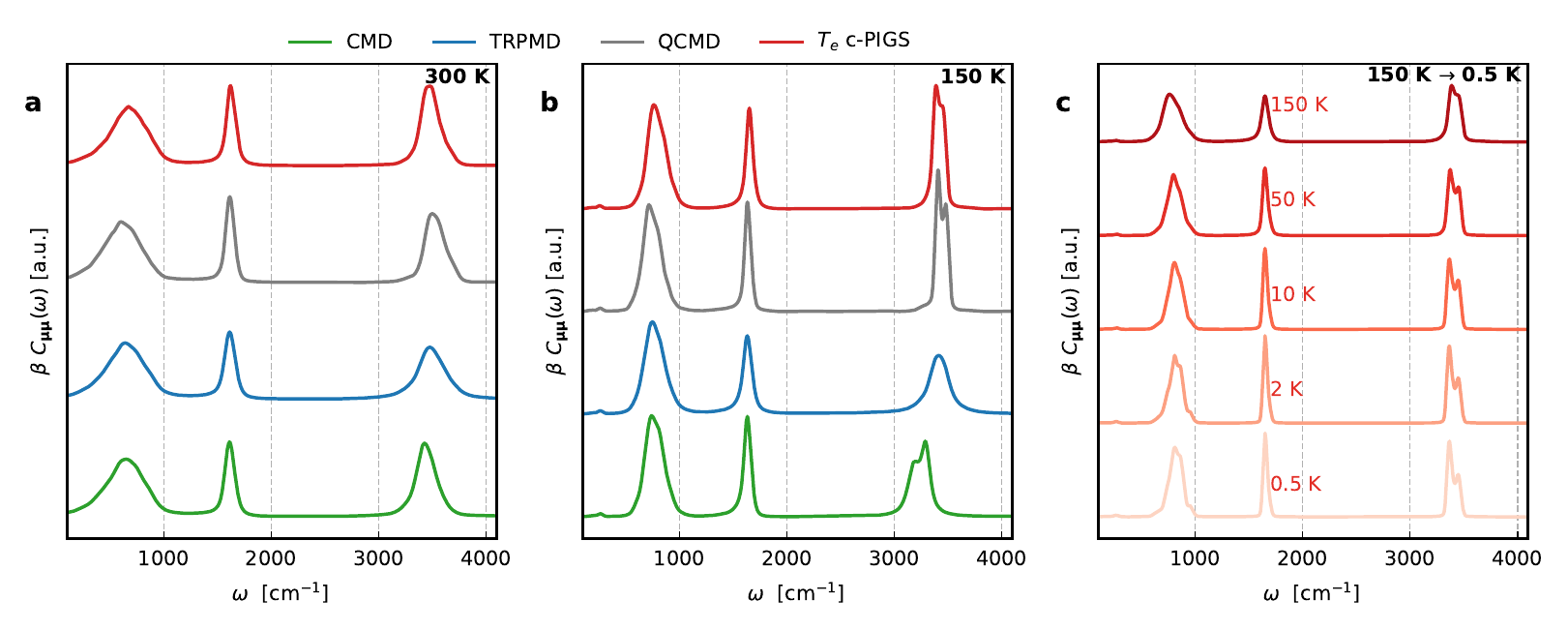}
    \caption{\scriptsize\textbf{Quantum dynamics of bulk water.} Comparison of the vibrational spectrum of (a) liquid water at 300\,K and (b) hexagonal ice at 150\,K, described by a q-TIP4P/f~\cite{habershon_competing_2009} water model, calculated using QCMD in black, the proposed \TePIGS{} approach using the centroid free energy surface calculated at 600 K in red, CMD in blue, and TRPMD in green. Panel (c) displays the \TePIGS{} IR spectrum predictions for hexagonal ice from 150 K down to 0.5 K.}
    \label{fig:h2obulk}
\end{figure*}

We demonstrate the capabilities of the \TePIGS{} approach by applying it to the IR spectrum of a water molecule -- a challenging system that exhibits a strong red-shift of the stretching modes due to zero-point motion and a fine temperature-dependent splitting of the vibrational peaks due to the coupling of rotations and vibrations.
We study the IR spectrum at 150\,K, 300\,K, and 600\,K using the well-known Partridge-Schwenke model~\cite{partridge_determination_1997}, which exhibits spectroscopic accuracy, and compare with experimental and numerically exact results~\cite{gao_phase_2018}.
We also compare with state-of-the-art approaches like CMD and thermostatted ring polymer molecular dynamics (TRPMD)~\cite{rossi_how_2014}. \\ 

The water molecule resides in its vibrational ground state at least up to 600\,K, but exhibits a sensitive dependence of the ro-vibrational splitting of the modes with temperature, as seen from the numerically exact IR spectra~\cite{trenins_path-integral_2019} in Fig.~\ref{fig:h2omol}. 
The TRPMD approach largely captures the correct line positions of the (envelopes of) the stretching and the bending bands but is artificially broadened~\cite{rossi_how_2014}.
The CMD approach gives a good description of the full IR spectrum at 600\,K but is artificially red-shifted at 300\,K and 150\,K due to the onset of the curvature problem. 
This is confirmed by observing in Fig.~\ref{fig:h2omol} that the centroid distributions at 300\,K and 150\,K indicate a lower (unphysical) O--H bond length and misalignment with the physical radial distribution. 
Nevertheless, at 600\,K we do not observe any ``symptoms" of the curvature problem, and thus we use $T_\mathrm{e} = 600$\,K for the \TePIGS{} approach. \\

As shown in Fig.~\ref{fig:h2omol}, \TePIGS{} spectra at 600\,K are in excellent agreement with the exact and the independent CMD results, underlying the accuracy of the learned FES. 
More importantly, \TePIGS{} describes the fundamental frequencies and the rotational splittings of the stretching and the bending modes at 300\,K and 150\,K, a substantial improvement over state-of-the-art PI methods in terms of accuracy.
\rev{We note however that, like CMD, \TePIGS{} does not capture the subtle temperature-dependence of the relative intensities of the bending mode w.r.t to the stretching mode.}
\rev{This is largely an artefact of PI-based approximate quantum dynamics methods which lack quantum coherence in their dynamics}~\cite{benson_matsubara_2021} \rev{arising from the momenta not being drawn from their quantum Boltzmann distribution}~\cite{ple_anharmonic_2021}.
On the other hand, \TePIGS{} is computationally efficient since the FES is estimated from a 100\,ps PI trajectory with 8 replicas and a timestep of 0.5\,fs, suggesting computational gains of at least a factor of $500\times$, $1000\times$ and $2000\times$ at 600\,K, 300\,K and 150\,K, respectively, if one knows an appropriate $\Te{}$, compared to a 1\,ns long CMD simulations needed to converge the TCFs. 
\rev{The overall cost for identifying $\Te{}$, namely a set of 100\,ps PIMD simulations across 100\,K-600\,K, was at least two orders of magnitude lower than that of a CMD simulation needed to calculate the spectrum at 150\,K.}
\rev{Moreover, we expect this cost to be overestimated, due to the over-conservative nature of our benchmark.}
\rev{These could easily be made an order of magnitude less-expensive using out-of-the-box accelerated-sampling methods~\cite{markland_nuclear_2018} based on generalized Langevin equation thermostats~\cite{ceriotti_accelerating_2011, ceriotti_efficient_2012, mauger_nuclear_2021} and high-order splittings of the Boltzmann operator~\cite{perez_improving_2011,kapil_high_2016, kapil_modeling_2019, shepherd_efficient_2021} combined with replica-exchange~\cite{sugita_replica-exchange_1999} across 100\,K-600\,K.}
\rev{Other accelerated techniques like ring polymer contraction and multiple time stepping-\cite{markland_efficient_2008, tuckerman_reversible_1992} promise a classical computational-cost for the calculation of quantum dynamical properties~\cite{marsalek_quantum_2017, kapil_accurate_2016}, contingent to finding an inexpensive surrogate for the high-frequency modes of the system, however, they bear the inaccuracies of traditional path-integral methods that we address using the $\Te{}$ \textit{ansatz}.} \\
To showcase the absence of computational scaling with temperature, we calculate the IR spectrum of a water molecule down to 0.5\,K. 
Decrease in temperature expectedly reduces the rotational splitting of the vibrational modes, and below 10\,K, the system falls into its rotational ground state with well resolved peaks for the three vibrational modes. 
This temperature is in excellent agreement with the experimentally known rotational constants of a water molecule, i.e. 13-35\, K~\cite{hall_pure_1967}. 
Furthermore, the frequencies of the ground state vibrational bending mode matches the experimental / exact results~\cite{gao_phase_2018} up to 8\,\invcm{} -- the resolution of calculated spectra -- and those of the two stretching modes are expectedly blue-shifted by around 40\,\invcm{} in excellent agreement with the results obtained from Mastubara dynamics~\cite{trenins_mean-field_2018} -- the true reference for PI based dynamical methods.
We emphasize that the calculation of quantum mechanical spectrum at cryogenic temperatures by means of classical dynamics at this level of accuracy-to-cost ratio is unprecedented to our knowledge. \\

As a final test of our approach, we study the IR spectrum of condensed phase aqueous systems: bulk water at 300\,K and hexagonal ice at 150\,K. 
The presence of inter-molecule or crystal modes that couple with high-frequency modes should constitute a challenge for \TePIGS{} as it has been the case with previous PI approaches~\cite{willatt_approximating_2018, rossi_fine_2017}.
An additional challenge is that ice melts below the onset temperature of the curvature problem which could complicate the calculation of the FES at $\beta_\mathrm{e}$.
To circumvent this issue, we exploit the local nature of the FES (see Eq.~\ref{eq:pmf_ansatz}) and insights from deep inelastic neutron scattering experiments~\cite{burnham_origin_2006} that probe the quantum nuclear motion of atoms. 
These experiments suggest that the local potential felt by the nuclei due to quantum delocalization is short ranged~\cite{lin_displaced_2010} and sensitive/unique to their local environments~\cite{andreani_direct_2016}.
Given that Eq.~\ref{eq:pmf_ansatz} makes the fitted FES size-extensive and local environments of H and O atoms in ice are present in liquid water~\cite{gasparotto_recognizing_2018}, we conjecture that the $\Te{}$ FES of hexagonal ice can be constructed from a high-temperature simulation of liquid water. 
We estimate vibrational spectra using the q-TIP4P/f water potential~\cite{habershon_competing_2009} and a linear dipole moment surface.
Although this model doesn't exhibit ``experimental" or ``spectroscopic" accuracy, it has been extensively used for comparing the performance of various PI
~\cite{rossi_communication:_2014, kapil_inexpensive_2020, benson_which_2019, trenins_path-integral_2019} and wavefunction-based\cite{wang_ab_2011, gruenbaum_robustness_2013} methods and exhibits good agreement with the experimental spectra when combined with an appropriate dipole moment surface~\cite{liu_transferable_2016}. \\

As shown in Fig.~\ref{fig:h2obulk}, we obtain well resolved spectra for both liquid water at 300\,K and hexagonal ice at 150\,K using a $T_\mathrm{e} = 600$\,K $\kb{}$ FES fitted on a 10\,ps PI simulation of liquid water. 
Our results are in good agreement with CMD, TRPMD and QCMD for room-temperature liquid water, where the curvature problem is small~\cite{paesani_quantitative_2010}.
In the case of hexagonal ice at 150\,K, we see a quantitative and qualitative improvement on the description of the high frequency band with respect to CMD and TRPMD and an excellent agreement with QCMD~\cite{trenins_path-integral_2019}. 
In addition to the increased accuracy, our approach is over three orders of magnitude less expensive (as measured by the number of force evaluations) than state-of-the-art approaches at 300\,K and 150\,K.
Furthermore, the absence of temperature-dependent scaling behaviour of the cost, allows us to also compute the IR spectrum of hexagonal ice all the way down to 0.5 K. 
As expected, we observe that the system resides in its ground state and does not exhibit any line shifts of the high frequency modes.   \\

\rev{In summary, we propose a new approach, \TePIGS{} that accurately simulates the quantum dynamics of light nuclei at the cost of classical MD by combining PI quantum mechanics, machine learning and bottom up coarse-graining.}
\rev{This combination is both timely and significant as exemplified by other recent work(that we discovered out during the review process) that implement a \textit{fast} version of CMD~\cite{loose_centroid_2022,wu_learning_2022} using machine learning potentials}.     
We use a physically-motivated temperature elevation \textit{ansatz} that moves the system on the centroid FES obtained at a high temperature, and fitted efficiently by using machine-learning-based coarse-graining approaches~\cite{Wang2019,husic2020coarse,wang2020ensemble}.
The $\Te{}$ \textit{ansatz} is exact in the high-frequency harmonic limit where the vibrational mode essentially lives in its ground state, as well as for low-frequency modes for which the ring polymer distribution collapses on the centroid.
Furthermore, our study of condensed aqueous phases water shows that it also performs well for intermediate frequencies $\hbar\omega \sim \beta^{-1}$ , for instance, the librational / rotational modes.
These limits suggest that our approach could be useful for studying a wide range of systems with high-frequency modes that exhibiting weak coupling to rest of the system. 
We show that the \TePIGS{} FES is \emph{transferable} across temperatures and phases, but is also size extensive, meaning that it allows for further computational savings by learning the FES on a smaller system. 
We believe that our approach constitutes a substantial improvement in accuracy over routinely used state-of-the-art methods like CMD and TRPMD.
In addition, its low computational cost and the absence of its scaling behavior with temperature allows for accurate the IR spectra predictions even at cryogenic temperatures -- considered prohibitive with state-of-the-art methods. \\

\rev{The routine use of the $\Te{}$ PIGS approach on materials and chemical system will require a careful and thorough study on a diverse set of ``difficult" systems which could challenge the accuracy of the $\Te{}$ \textit{ansatz}. 
For instance, one could imagine highly fluxional molecules, system exhibiting a near-continuum of strongly coupled high-frequency modes, or dynamical processes dominated by ``rare" quantum tunnelling events, being testing cases probing the regimes in which the $\Te{}$ \textit{ansatz} is not exact or expected to be accurate. 
Similarly, the extension of $\Te{}$ PIGS to general systems described by first-principles electronic structure methods, will require development of a hierarchical framework leveraging recent advances in active-learning strategies for the development of accurate and reliable machine-learning PES, enhanced sampling of the quantum Boltzmann distribution using accelerated path-integral methods, and subsequently the calculation of the centroid FES at an elevated temperature. 
Additionally, these systems would also form an ideal test bed for understanding what range of elevated temperatures could be used to treat a frozen ground-state vibrational mode. 
Could a ``universal" range of $\Te{} \in [500,600]$\,K. be used for chemical systems, displaying high-frequency modes in the 3000 - 5000 cm$^{\text{-1}}$ regime, to completely avoid process of selecting an elevated temperature for a general system?
And finally, one could envisage the use of the PIGS approach for calculations of equilibrium properties. 
We plan to address all these directions in future studies. 
}
Overall, we believe that the simplicity, low cost and high accuracy of $\Te{}$ PIGS could open up prospects for routine modeling of quantum-vibrational spectra of general systems and direct comparisons with experiments, often performed at low temperatures~\cite{verma_infrared_2019}. \\

\section*{Acknowledgements}
We thank Start Althorpe, George Trenins, Christoph Schran, Angelos Michaelides, and members of the Clementi's group at FU for insightful discussions and comments on the manuscript. 
We also thank George Trenins for sharing QCMD results. V.K. acknowledges funding from the Swiss National Science Foundation (SNSF) under Project $\text{P2ELP2}\_\text{191678}$ and the Ernest Oppenheimer Fund,  allocation of CPU hours by CSCS under Project ID s1000, and support from Churchill College, University of Cambridge.
C.C. acknowledges funding from the Deutsche Forschungsgemeinschaft DFG (SFB/TRR 186, Project A12; SFB 1114, Projects B03 and A04; SFB 1078, Project C7; and RTG 2433, Project Q05), the National Science Foundation (CHE-1738990, CHE-1900374, and PHY-2019745), and the Einstein Foundation Berlin (Project 0420815101).
F.N. acknowledges funding from the Deutsche Forschungsgemeinschaft DFG (SFB 1114, Projects A04 and B08), The Berlin Mathematics center MATH+ (Projects AA1-6, AA2-8), The Berlin Institute for the Foundations of Learning Data (BIFOLD) and the European Commission (ERC CoG 772230).
F.M. acknowledges support from the SNSF under the Postdoc.Mobility fellowship P500PT\_203124 and from the Physics department at FU Berlin for the computational time.

\section*{Source data}
All the raw data and code required to reproduce the results, and scripts and raw data needed to make the figures is made available in the online repository:  {\href{https://github.com/venkatkapil24/pigs-methodology}{\texttt{venkatkapil24/pigs-methodology}}}.

\end{document}


\title{Supporting Information: Quantum dynamics using path integral coarse-graining}

\author{F\'elix Musil}
\affiliation{Department of Physics, Freie Universit\"at Berlin, Arnimallee 12, 14195 Berlin, Germany}

\author{Iryna Zaporozhets}
\affiliation{Department of Physics, Freie Universit\"at Berlin, Arnimallee 12, 14195 Berlin, Germany}
\affiliation{Department of Chemistry, Rice University, Houston, Texas 77005, United States}
\affiliation{Center for Theoretical Biological Physics, Rice University, Houston, Texas 77005, United States}

\author{Frank No\'e}
\affiliation{Department of Physics, Freie Universit\"at Berlin, Arnimallee 12, 14195 Berlin, Germany}
\affiliation{Center for Theoretical Biological Physics, Rice University, Houston, Texas 77005, United States}
\affiliation{Department of Chemistry, Rice University, Houston, Texas 77005, United States}
\affiliation{Microsoft Research, Cambridge, UK}

\author{Cecilia Clementi}
\email{cecilia.clementi@fu-berlin.de}
\affiliation{Department of Physics, Freie Universit\"at Berlin, Arnimallee 12, 14195 Berlin, Germany}
\affiliation{Center for Theoretical Biological Physics, Rice University, Houston, Texas 77005, United States}
\affiliation{Department of Chemistry, Rice University, Houston, Texas 77005, United States}

\author{Venkat Kapil}
\email{vk380@cam.ac.uk}
\affiliation{Yusuf Hamied Department of Chemistry,  University of Cambridge,  Lensfield Road,  Cambridge,  CB2 1EW,UK}

\maketitle

\section{Computational details}

\subsection{DVR calculations}

\noindent The 2D radial Morse potential mimicking the O--H bond, as described in the main text is defined as
\begin{align}
    \hat{H} = (2 \mu)^{-1} \left[\hat{p}_x^2 + \hat{p}_y^2\right] + D \left[1 - e^{-\alpha \left(\sqrt{\hat{q}_x^2 + \hat{q}_y^2} - r_0\right)}\right]^2, 
    \label{eq:si_2d_morse}
\end{align}
with $\mu = 1741.1$ a.u. the reduced mass of O and H atoms, $D = 0.18748$ a.u. the bond dissociation energy, $r_0 = 1.8324$ a.u. is the bond length parameter and $\alpha = 1.1605$ a.u. a parameter describing the the anharmonicity along the bond. 
We solved the time-independent Schr\"odinger's equation using sinc-function discrete variable representation~\cite{colbert_novel_1992} implemented in Nuclear Solver~\cite{graen_nusol_2016}. 
The potential was solved on a grid with dimensions $x$ = [-6.0, 6.0] a.u. and $y$ = [-6.0, 6.0] a.u. using a uniform grid with 400 points in each direction. 
These parameters yield converged values for at least the first 32 eigenvalues.

\subsection{PIMD simulations}

\noindent We performed PIMD simulations of all systems in the canonical ensemble using the PILE-L and PILE-G thermostats~\cite{ceriotti_efficient_2012} of time constants 100\,fs for building the datasets used to train the machine learning models. 
We used a timestep of 0.5\,fs and integrated the PI equations of motion using a BAOAB splitting~\cite{kapil_modeling_2019} of the Liuoville operator. 
We sampled the centroid positions and forces every 10\,fs. Simulations were run for 100\,ps at temperatures $\{100\,\textrm{K}, 200\,\textrm{K}, \dots, 900\,\textrm{K}, 1000\,\textrm{K}\}$ using 64 replicas at all temperatures using the \texttt{i-PI} code~\cite{kapil_i-pi_2019}.

\subsection{ML model}

We model the atomic potential of mean force, $A_a$, by representing the atomic environment with the normalized SOAP power spectrum~\cite{Bartok2013,Musil2021} and pass these 3-body features to multi-layer perceptron with $3$ hidden layers of width $[400, 200, 200]$ and the hyperbolic tangent activation function. 
The SOAP features are defined by:
\begin{equation}
    p^{(i)}_{a_1a_2n_1n_2l} = \sum_m (-1)^m c^{(i)}_{a_1n_1lm} c^{(i)}_{a_2n_2l (-m)},
\end{equation}
where 
\begin{equation}
    c^{(i)}_{anlm} = \sum_{(j,b)\in i} \delta_{ab} R_{nl}(r_{ij}) Y_l^m(\hat{\mathbf{r}}_{ij}) f_c(r_{ij})
    \label{eq:spherical_expansion}
\end{equation}
are the spherical expansion coefficients associated with atom $i$, the sum runs over the neighbors $j$ of species $b$ of atom $i$, $R_{nl}$ is a radial basis function, $Y_m^l$ are spherical harmonics functions, $f_c$ is a smooth cutoff function of radius $r_c$, $\mathbf{r}_{ij}=\mathbf{r}_{j} - \mathbf{r}_{i}$ is the separation vector going from the central atom $i$ to the neighbor atom $j$, $r_{ij}=\|\mathbf{r}_{ij}\|$ is its length, and $\hat{\mathbf{r}}_{ij}= \mathbf{r}_{ij} / r_{ij}$ is its direction.
The basis expansion in \cref{eq:spherical_expansion} is truncated up to $n<12$ radial channels and $l<=6$ angular channels.
Furthermore, $R_n$ is the splined GTO radial integral\cite{Musil2021} with $\sigma=0.3$, the cutoff radius is $r_c=3~\AA$, and the number of SOAP features is reduced by using a lossless compression scheme~\cite{Darby2021}. 
The parameters of the multi layer perceptron are optimized using the LAMB~\cite{You2019} ($l_r =0.001, \beta= (0.9, 0.999)$) and stochastic weight averaging~\cite{Izmailov2018}.

\subsection{CMD simulations}

\noindent We performed partially adiabatic CMD simulations~\cite{hone_comparative_2006} of all systems in the canonical ensemble using a weak PILE-G thermostat~\cite{ceriotti_efficient_2012} of time constants 1000\,fs and a Parrinello-Rahman mass matrix~\cite{parrinello_study_1984} to rescale the non-centroid frequencies to 13000\,\invcm{}.
We used a timestep of 0.01\,fs and integrated the PI equations of motion using an OBABO splitting~\cite{kapil_modeling_2019} of the Liouville operator. 
We sampled the centroid positions every 0.25\,fs. 
Simulations were run for 100\,ps at temperatures 150\,K, 300\,K and 600\,K using 64, 32, and 160 replicas, respectively, using the \texttt{i-PI} code~\cite{kapil_i-pi_2019}.

\subsection{MD / TRPMD simulations and PIGS}

\noindent We performed partially adiabatic TRPMD simulations~\cite{hone_comparative_2006} of all systems in the canonical ensemble using a PILE-G thermostat~\cite{ceriotti_efficient_2012} of time constants 100\,fs and PILE-$\lambda$=0.5.
We used a timestep of 0.25\,fs and integrated the PI equations of motion using an OBABO splitting~\cite{kapil_modeling_2019} of the Liuoville operator. 
We sampled the bead positions every 0.25\,fs. 
Simulations were run for 100\,ps at temperatures 150\,K, 300\,K and 600\,K using 64, 32, and 160 replicas, respectively, using the \texttt{i-PI} code~\cite{kapil_i-pi_2019}. \\
MD simulations and PIGS were performed by setting the number of replicas to one. 
Note that the simulations used to compute the IR spectra of water have been performed in two steps in order to sample the system sufficiently. 
First an NVT simulation with the Langevin thermostat was performed from which 100 configurations were sampled randomly to start NVT simulations with a weak stochastic velocity rescaling thermostat~\cite{bussi_canonical_2007} simulations -- known to improve sampling without significantly perturbing the dynamics of the system~\cite{ceriotti_efficient_2012}.

\section{The temperature elevation \textit{ansatz}}

\subsection{Ground state quantum time correlation function}

\noindent The standard time correlation function (TCF) of operators $\hat{A}$ and $\hat{B}$ for a Hamiltonian $\hat{H}$ at an inverse temperature $\beta = \kb{} T $ is 

\begin{equation}
c_{\hat{A} \hat{B}}(t) = Z^{-1}~\mathrm{Tr}\left[e^{-\beta \hat{H}}~\hat{A}~ e^{i \hbar^{-1}\hat{H}~t}~ \hat{B}~ e^{-i \hbar^{-1}\hat{H}~t} \right],
\label{eq:std_TCF}
\end{equation}
where $Z$ is the canonical partition function of the system. 
One can rewrite Eq.~\ref{eq:std_TCF} by computing the trace in the basis of the eigenstates of $H$, and inserting a resolution of identity
\begin{equation}
c_{\hat{\mathbf{\mu}} \hat{\mathbf{\mu}}}(t) = Z^{-1}~ \sum_{j,k} e^{-\beta E_j}<j|{\hat{A}}| k> <k|{\hat{B}}| j> ~ e^{i \hbar^{-1}(E_k - E_j)~t}
\end{equation}
with $\hat{H} \left|i\right> = E_i \left|i\right>$. The Fourier transform of the TCF is
\begin{equation}
{C}_{\hat{A} \hat{B}}(\omega) = Z^{-1}~ \sum_{j,k} e^{-\beta E_j}<j|{\hat{A}}| k> <k|{\hat{B}}| j> ~ \delta\left(\omega -  \hbar^{-1}(E_k - E_j)\right).
\label{eq:std_FTCF}
\end{equation}
For a system in the ground state, i.e., $\hbar \omega >> \beta^{-1}$, and thus the Boltzmann weight of the excited vibrational states can be ignored. Eq.~\ref{eq:std_FTCF} can be rewritten as a temperature-independent sum over matrix elements of the dipole operator:
\begin{equation}
{C}_{\hat{A} \hat{B}}(\omega) \approx \sum_{k} <j|{\hat{A}}| k> <k|{\hat{B}}| j> ~ \delta\left(\omega -  \hbar^{-1}(E_k - E_j)\right)
\end{equation}
The corresponding Kubo-transformed TCF $\tilde{C}_{\hat{A} \hat{B}}(\omega)$ is linked by the harmonic detailed balance relation $C_{\hat{A} \hat{B}}(\omega) = \tilde{C}_{\hat{A} \hat{B}}(\omega) \frac{\beta\hbar\omega}{1 - e^{-\beta\hbar\omega}}$. For a system in the ground state we apply the $\beta \hbar\omega \to \infty$ limit and arrive at 
\begin{equation}
\beta \tilde{C}_{\hat{A} \hat{B}}(\omega) \approx ({\hbar\omega})^{-1} \sum_{k} <j|{\hat{A}}| k> <k|{\hat{B}}| j> ~ \delta\left(\omega -  \hbar^{-1}(E_k - E_j)\right).
\label{eq:kubo_FTCF}
\end{equation}

\subsection{Choice of $T_\mathrm{e}$}

\noindent The choice of $T_\mathrm{e}$ depends on the anharmonicity of the system as $T_\mathrm{e}$ should be large enough to alleviate the curvature problem but small enough for the harmonic approximation to be valid.
For instance, for a harmonic system, an arbitrarily large $T_\mathrm{e}$ is a suitable as the classical and the quantum \textit{fundamental} transition  frequencies are equal.
On the other hand, for anharmonic systems, there is a need to select a low enough $T_\mathrm{e}$ as 
the $T_\mathrm{e} \to \infty$ limit yields the (oft blue-shifted) classical line position. 
As discussed in the main text, an easy and robust approach for selecting a $T_\mathrm{e}$ is by comparing the alignment of the centroid and the physical probability distributions. \\

\begin{figure*}[!h]
    \centering
    \includegraphics[width=\textwidth]{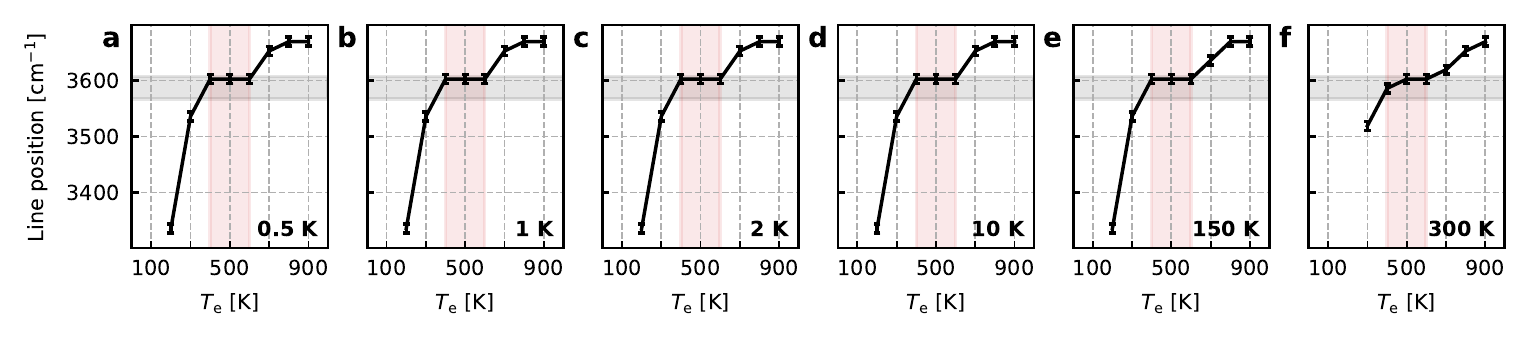}
    \caption{\scriptsize\textbf{Dependence of the line position of a 2D Morse oscillator on $T_\textrm{e}$.} The black shaded region shows the range between frequencies obtained by DVR (numerically exact) and Matsubara dynamics~\cite{trenins_path-integral_2019}. The red shaded region is used to indicate the range of suitable $T_\textrm{e}$ values.}
    \label{fig:te}
\end{figure*}

\noindent As shown in Fig.~\ref{fig:te}, the $T_\mathrm{e}$ dependence of the fundamental transition frequency displays three distinct regimes. 
The curvature problem is attenuated for $T_\mathrm{e} <$  400\,K and beyond 600\,K, anharmonic effects lead to a systematic blue shift to the classical transition frequency at $T_\mathrm{e} \to \infty$. 
We observe a \textit{goldilocks scenario} in the temperature range, 400\,K to 600\,K, where the curvature problem is cured and anharmonic effects are small; any $T_\mathrm{e}$ in this range of temperature should be a valid choice.

\newpage